# Terahertz channel modeling based on surface sensing characteristics

Jiayuan Cui[1], Da Li[1], Jiabiao Zhao[1], Jiacheng Liu[1], Guohao Liu[1,2], Xiangkun He[3], Yue Su[1,2], Fei Song[4], Peian Li[1], Jianjun Ma[1,2,5]

[1]School of Integrated Circuits and Electronics, Beijing Institute of Technology, Beijing 100081, China

[2]Tangshan Research Institute, BIT, Tangshan, Hebei 063099, China

[3]Department of Bioengineering, Faculty of Engineering, Imperial College London, London SW7 2AZ, UK

[4]School of Electronic and Information Engineering, Beijing Jiaotong University, Beijing 100044, China

[5]Beijing Key Laboratory of Millimeter and Terahertz Wave Technology, Beijing 100081, China

**Abstract** - The dielectric properties of environmental surfaces, including walls, floors and the ground, etc., play a crucial role in shaping the accuracy of terahertz (THz) channel modeling, thereby directly impacting the effectiveness of communication systems. Traditionally, acquiring these properties has relied on methods such as terahertz time-domain spectroscopy (THz-TDS) or vector network analyzers (VNA), demanding rigorous sample preparation and entailing a significant expenditure of time. However, such measurements are not always feasible, particularly in novel and uncharacterized scenarios. In this work, we propose a new approach for channel modeling that leverages the inherent sensing capabilities of THz channels, specifically by obtaining channel measurement data through the analysis of refractive indices. By comparing the results obtained through channel sensing with that derived from THz-TDS measurements, we demonstrate the its ability to yield dependable surface property information. Integrating it into a ray-tracing algorithm for channel modeling in both a miniaturized cityscape scenario and an indoor environment, the results show consistency with experimental measurements, thereby validating its effectiveness in real-world settings.

## I. Introduction

The advent of innovative technologies has prompted an exponential growth in wireless data transmission capabilities [1]. It is expected that the data capacity and user density requirements of future 6G wireless networks will dwarf those of the 5G predecessors by orders of magnitude, aiming for data rates in the range of terabits per second (Tbps) [2]. However, current spectrum allocations fall short of supporting these high data rates, necessitating the exploration of higher frequency bands, such as the terahertz (THz) frequency [3]. THz band is one of the promising candidates for high-speed data transmission owing to its large bandwidth. For an example, the 275-450 GHz band has been identified for land mobile and fixed service applications at the 2019 World Radio communication Conference (WRC-19) [4]. Meanwhile, the THz band is also used in wireless cognition, channel sensing and imaging technology, due to its shorter wavelength, which allows for high-resolution imaging and precise sensing capabilities [5-7]. In order to progress the application of THz frequency technology, accurate channel modeling becomes necessary for the design and evaluation of the systems and protocols.

The precision of channel models relies critically on the accurate representation of the dielectric properties of environmental materials. Commonly, the effort for reliable THz channel modelling has been challenged by the variability and unknown nature of these dielectric properties, especially when propagating in uncharacterized scenarios. Recent endeavors have aimed to address these challenges, embarking on comprehensive studies and measurements of THz channels in many kinds of scenarios. Notable efforts include outdoor citywide channel measurements in the 140-141 GHz band, yielding significant insights into capacity [8-10], and bidirectional THz channel measurements in microcellular environments, comparing parameters like path loss models and delay spread [11-13]. Additionally, angle-resolved measurements at 300 GHz have been conducted outdoors [14], alongside indoor measurements at the 350 GHz and 650 GHz frequencies [15], with findings on path loss indices and consistency across frequencies in indoor settings [16-18]. Investigations have also been carried out on time delay and angle propagation in various indoor scenarios [19, 20], as well as on comparing radio propagation characteristics in malls and office environments at 28 GHz and 140 GHz [21, 22]. Reflective and absorptive materials' impacts on channel characteristics have been examined through bidirectional channel measurements in diverse scenarios [23].

For modeling, innovations have emerged, such as the development of a THz D2D indoor geometric random channel model [24] and a three-dimensional model to capture propagation parameters for indoor communications [25, 26]. A ray-tracing algorithm has been proposed for a diffuse scattering model applicable in THz communications [27], along with a hybrid model combining ray tracing and statistical methods for indoor THz communications [28, 29]. Recently, dielectric constants of various building materials were measured using THz time-domain spectroscopy (THz-TDS) systems to inform the reflectance and transmittance calculations for indoor propagation modeling at the 300 GHz frequency [30]. However, when facing the dynamics of channel movement and propagation within unfamiliar and uncharacterized environments, the endeavor encounters significant hurdles [31], with the dielectric properties difficult to obtain accurately and timely. The accurate and timely acquisition of dielectric properties, essential for precise channel modeling, becomes markedly challenging. This difficulty is amplified in scenarios where the channel characteristics are in constant variation, requiring rapid adaptation and model updating to accurately reflect the real-world conditions.

In this work, we introduce an approach to overcome this obstacle by leveraging the surface sensing capabilities of THz channels. By sensing the surface properties directly, this method proposes to improve the accuracy of THz channel models, integrating these findings with the aforementioned ray-tracing algorithms to analyze channel characteristics comprehensively. The rest of this article is organized as follows. In Section II, we describe a miniaturized cityscape scenario and experimental setup. Then the reflectance characteristics and dielectric constants of the wall and window materials of the model building are measured by channel characteristic sensing. Finally, the differences between the theoretical and experimental values are analyzed. In Section III, we measure THz channels and characterize them based on ray-tracing algorithms. At the same time, a THz channel model is established based on the channel sensing characteristics, and the actual measurement data is compared with the modeling results. In Section IV, we further verify the proposed method in a real indoor scenario, and discuss the differences between the model and the real communication scenario to verify the effectiveness of our proposed method. Finally, the paper is summarized in Section V.

## II. Channel measurement setup and channel sensing

For channel measurement, we build a miniaturized cityscape scenario in our laboratory (Fig. 1), instead of taking measurements in an actual city environment. This is crucial for isolating specific factors and understanding their effects on the THz channel without the interference of unpredictable variables that would be present in a real-city scenario, such as weather conditions, unexpected obstacles, and human activity [32, 33]. For the measurement setup, there is a Ceyear 1465D signal generator, capable of producing baseband signals across a spectrum from 100 kHz to 20 GHz. These signals undergo up-conversion to the desired THz frequency range (110-170 GHz) by a Ceyear 82406B frequency multiplier module, achieving a multiplication factor of ×12. The generated continuous wave (CW) signals are then launched out via a horn antenna (HD-1400SGAH25), while the receiving end is equipped with a matching horn antenna to forward the incoming signals towards a Ceyear 71718 power sensor for direct detection. For operations within the higher frequency domain (220-325 GHz), our configuration includes a Ceyear 82406D frequency multiplier, enhancing signals by a factor of ×18, in cooperation with a Ceyear 89901S horn antenna and a teflon lens (focal length 10 cm) to direct the beams. This assembly ensures beam gains of 25 dBi at 140 GHz and 10 dBi at 240 GHz, with the entire apparatus positioned 40 cm above the experimental platform to clear the first Fresnel zone. Notably, the antennas maintain a constant elevation angle of 0°, with the channel's beamwidth measured at 18° for the 140 GHz channel and approximately 4° for the 240 GHz channel. A laptop is employed for setup controlling and data acquisition, managing both the signal generator and power meter at a data recording frequency of 7 Hz.

Both the transmitter (Tx) and receiver (Rx) units are mounted on separate rotary tables, allowing precise control over their orientation relative to each other and the miniaturized cityscape environment. This setup enables detailed measurement of the angle-dependent power variation of the reflected signal, which is crucial for obtaining accurate refractive indices of the surface materials. The rotary tables are capable of rotating from -30° to +30°, which allows the Tx and Rx to be oriented at various angles to each other, spanning a total arc of 60°. The rotation step of 1° provides a high-resolution dataset of how the channel characteristics vary with changes in the angle of incidence and reception. The Tx and Rx are positioned on opposite sides of the cityscape scenario, as shown in Fig. 1(a). This setup is used to emulate the transmission and reception of THz channels in a city environment, where buildings affect channel propagation. The model allows for the study of reflection off building surfaces and validate our method for channel sensing. In this miniaturized scenario, we focus on the reflection of two specific types of surfaces: walls and windows. The walls in the scenario are emulated using a type of flat and smooth plastic materials. The choice of plastic is likely due to its uniformity, ease of shaping and controlling dimensions, and predictable interaction with THz waves. While plastics allow significant penetration at THz frequencies, our measurements focused on the reflected components, without considering the diffraction over the building edges and absorption by the materials. This is to be consistent with actual construction materials, which absorb a significant portion of the THz waves and minimize the impact of penetration components on reflection measurements [37]. The exact thickness of the plastic used to represent the walls is specified as 1.889 mm by a calliper, indicating a placeholder for the actual measurement. Similarly, the refractive index, a critical parameter for understanding how THz waves are transmitted and reflected by the material, is measured as 1.733 at 140 GHz and 1.664 at 240 GHz. Representing window glass, another common city surface, is a different type of smooth plastic. This choice suggests an intention to mimic the transmission and reflection characteristics of glass at THz frequencies closely. Again, the thickness is marked as 0.239 mm, and the measured refractive index as 1.575 at 140 GHz and 1.561 at 240 GHz. These exact refractive index values are measured using a THz time-domain spectroscopy (THz-TDS) system (T-SPEC 800), which has been considered as a highly effective method for characterizing material's optical properties in the THz range.

Acknowledging the complexity associated with the unknown refractive indices of various surfaces, a pivotal limitation emerges in the precise modeling and prediction of THz channel behaviors. To deal with this obstacle, we propose an approach aimed at obtaining these properties through an examination of the reflection characteristics perceived by THz channels upon their interaction with diverse surfaces. The foundational principle of this methodology is illustrated in Fig. 2(a), where the channel traverses air (medium 1) before being reflected at the boundary between medium 1 and the surface (medium 2). This reflection process is mathematically described by the Fresnel equations,

$$\rho = \frac{\rho_{12} + \rho_{23} \cdot e^{-2 \cdot \lambda_2 \cdot d \cdot \cos\theta_2}}{1 + \rho_{12} \cdot \rho_{23} \cdot e^{-2 \cdot \lambda_2 \cdot d \cdot \cos\theta_2}}, \quad (1)$$

which facilitate the derivation of the dielectric properties of the encountered surfaces [34]. In this equation, $\theta_1$, $\theta_2$ and $\theta_3$ denote the angle of incidence in medium 1, 2, and 3, respectively. $\lambda_2$ denotes the THz wave's wavelength in medium 2, and $d$ specifies the thickness of this medium. The reflection coefficients at the interface between mediums 1 and 2 ($\rho_{12}$) and between mediums 2 and 3 ($\rho_{23}$), as shown in Fig. 2(a), and can be expressed as by Fresnel equation:

$$\rho_{12} = \frac{\sqrt{\varepsilon_1} \cdot \cos\theta_1 - \sqrt{\varepsilon_2} \cdot \cos\theta_2}{\sqrt{\varepsilon_1} \cdot \cos\theta_1 + \sqrt{\varepsilon_2} \cdot \cos\theta_2} \tag{2}$$

$$\rho_{23} = \frac{\sqrt{\varepsilon_2} \cdot \cos\theta_2 - \sqrt{\varepsilon_3} \cdot \cos\theta_3}{\sqrt{\varepsilon_2} \cdot \cos\theta_2 + \sqrt{\varepsilon_3} \cdot \cos\theta_3} \tag{3}$$

with the parameters $\varepsilon_1$, $\varepsilon_2$, $\varepsilon_3$ denoting the dielectric constants of mediums 1 (air), 2 (surface material), and 3 (air), respectively. Based on the assumption of smooth surfaces [35] in the theoretical model in Eq. (1) with multiple reflection components considered, we conduct perceptions, subsequently comparing that with our empirical data (by the setup in Fig. 1) as shown in Fig. 2(b) and (c). This comparison is achieved by orienting a THz channel toward the interface and then measuring the angle-dependent power variation of the reflected signal. The objective herein is to ascertain the accuracy of our perceptions based on Eq. (1). It can be seen that there is a difference between the calculated and measured reflectivity values, and it reduces to negligible at certain incident angles. This affirms the efficacy of our proposed method in accurately obtaining the dielectric properties of surfaces by sensing and analyzing their reflective behavior. Additionally, our method uses Eq. (1) to apply the measured signal reflectivity and obtain the dielectric constant of surface 2, assuming the dielectric constants of surfaces 1 and 3 are known. This process has been clarified and supplemented in the corresponding sections of the manuscript. For further confirmation, we compare the perceived values with the measured data at 140 GHz and 240 GHz by the THz-TDS as shown in Table I, which demonstrates that the average deviation from our predictions is below 6%. If the number of reflecting layers is three or more, the reflection coefficient can be calculated using $\rho_{ij} = \frac{\sqrt{\varepsilon_i} \cdot cos\theta_i - \sqrt{\varepsilon_j} \cdot cos\theta_j}{\sqrt{\varepsilon_i} \cdot cos\theta_i + \sqrt{\varepsilon_j} \cdot cos\theta_j}$.

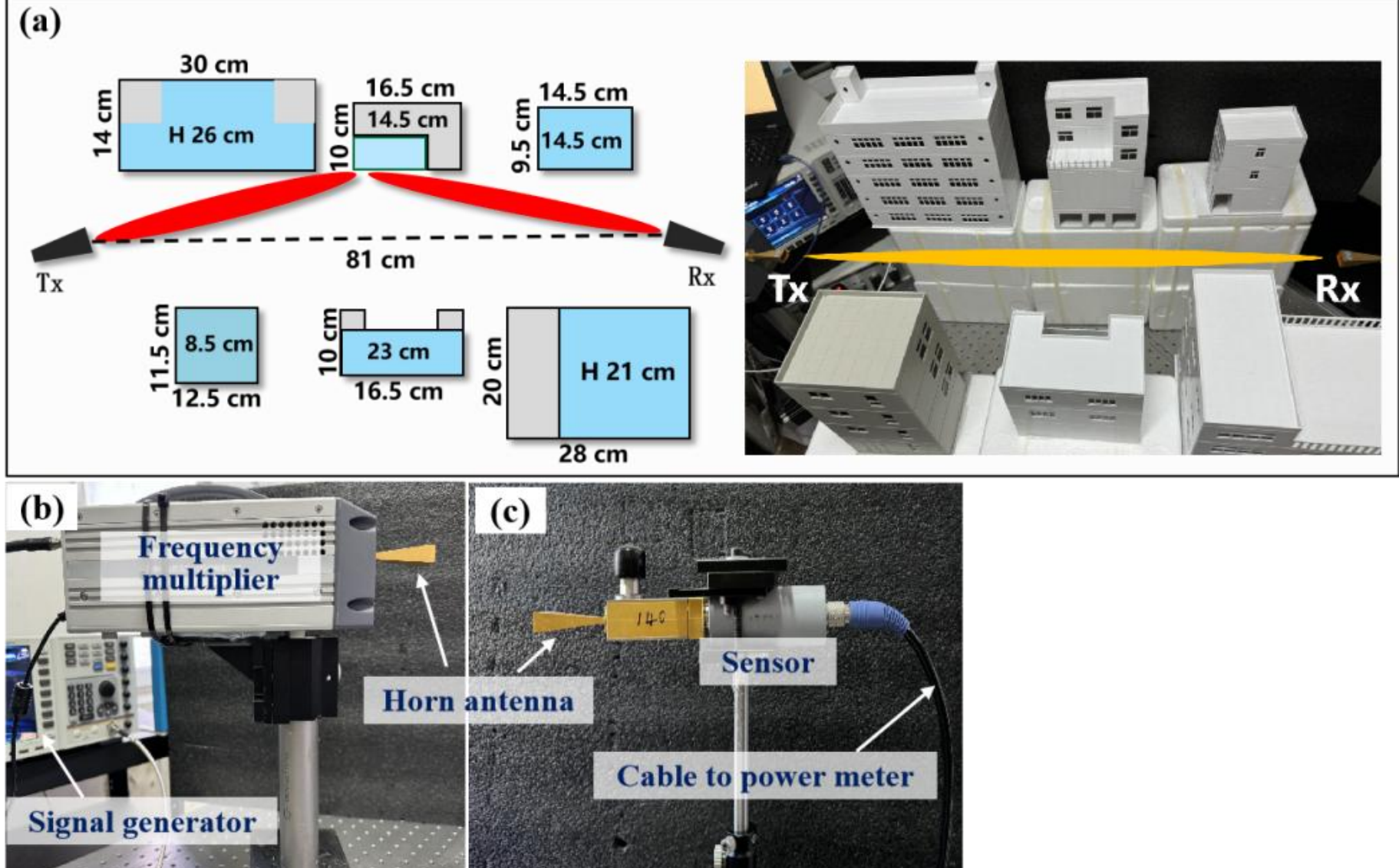

**Figure 1**. THz channel measurement setup. (a) Indoor channel measurement setup simulating an outdoor city environment through the implementation of a miniaturized cityscape model; (b) Transmitter hardware; (c) Receiver hardware.

**Table I** Perceived (by channel sensing) and measured (by THz TDS) refractive index of the surface materials.

| **Surface** | **Wall** | **Window** |
|---|---|---|
| Perceived @ 140 GHz | 1.681 | 1.619 |
| Measured @ 140 GHz | 1.733 | 1.575 |
| Error @ 140 GHz | 5.2 % | 4.4 % |
| Perceived @ 240 GHz | 1.672 | 1.607 |

| Measured @ 240 GHz | 1.664 | 1.561 |
| --- | --- | --- |
| Error @ 240 GHz | 0.8 % | 4.6 % |

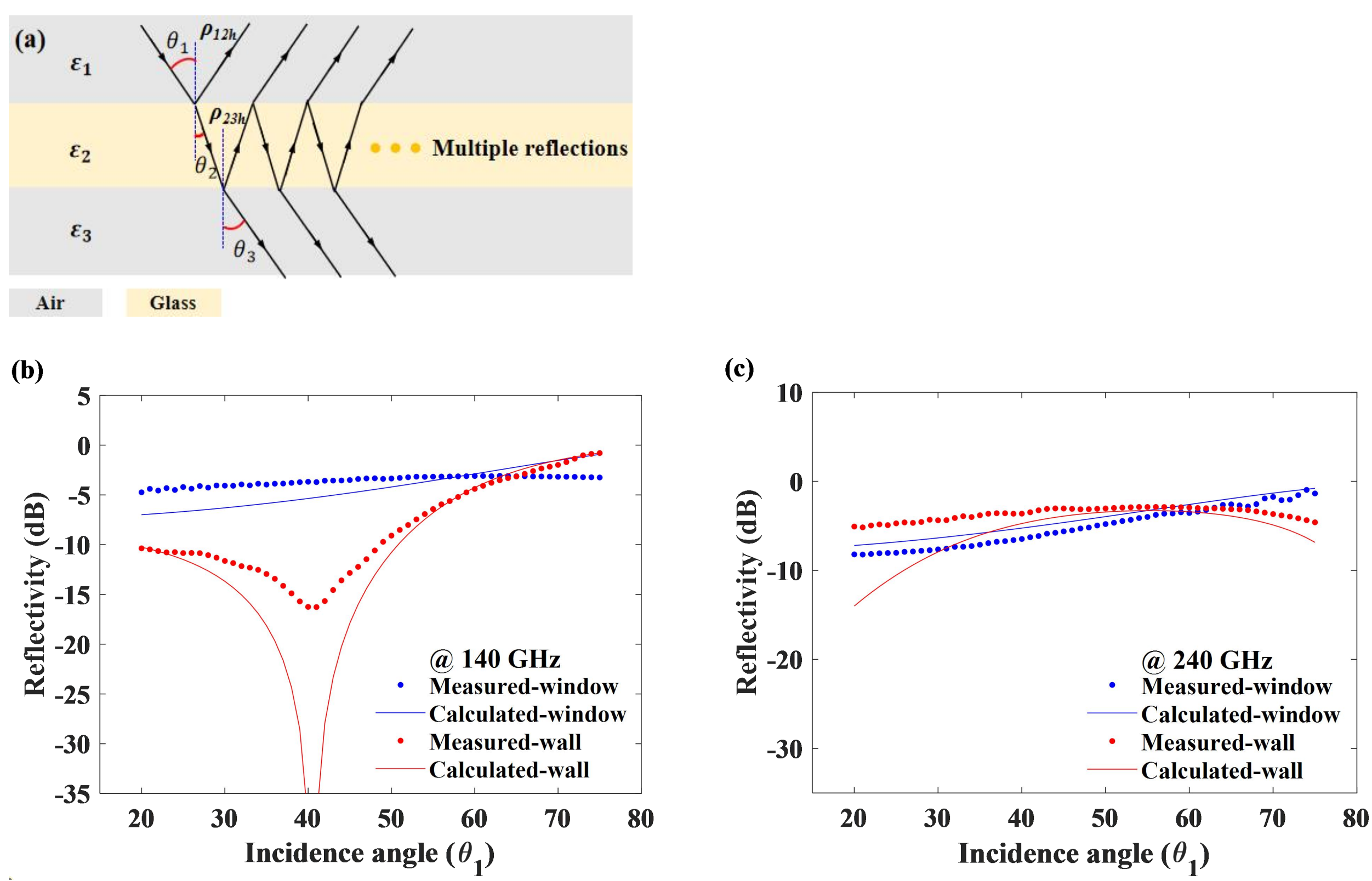

**Figure 2.** Measured and predicted reflection characteristics by wall and glass surfaces. (a) Illustration on reflection mechanism by a layer of dielectric medium (wall or glass) in air; (b) Reflection of 140 GHz channel by wall and window surfaces; (c) Reflection of 240 GHz channel by wall and window surfaces.

## III. Channel measurement and modeling in miniaturized cityscape scenario

In order to explore the efficiency of the perceived dielectric properties of building surfaces, we perform channel measurements on the miniaturized cityscape scenario, and establish a THz channel model based on the measured and perceived material properties above. The ray tracing algorithm, a computational technique widely recognized for its efficacy in simulating THz channel propagation through complex environments, meticulously models the interaction of THz waves with various obstacles and surfaces [16, 28]. Here, our implementation of the ray tracing algorithm intricately simulates the trajectory of THz channels launching from the transmitter, accounting for the phenomena of reflection, diffraction, and scattering as these rays encounter cityscape structures and indoor elements to analyze the characteristics of the THz channel. This approach allows for a comprehensive analysis of both LOS and NLOS components, critical for THz communications. Settings within the algorithm have been meticulously calibrated to align with the physical realities of our study environments, incorporating the specific geometries of miniaturized cityscape and indoor settings, the reflective properties of materials at THz frequencies, and the precise positions of the Tx and Rx units. Such detailed modeling ensures that the spatial power distribution maps generated offer a close approximation to actual THz channel behavior. During the modeling, the path delay and channel impulse responses (CIR) is recorded. Based on the ray tracing modeling results, the statistics of the THz channel in the miniaturized cityscape scenario are characterized in terms of path delay distribution for the LOS and NLOS cases while the cumulative distribution function (CDF) is made based on the signal delay. Moreover, the multi-path components (MPCs) are thoroughly investigated to reveal the characteristics of the THz channel.

We perform measurements on the miniaturized cityscape THz channel by the experimental method described in Fig. 1. When both the Tx and Rx are oriented towards the same side of the buildings, the channel is reflected by the building surface and generates NLOS; when Tx and Rx are directly facing each other, the channel is transmitted directly from Tx to Rx, generating a LOS channel. We process the measured data to make a spatial power distribution map (Fig. 3) by taking the azimuthal change of Tx as the horizontal coordinate and the azimuthal change of Rx as the vertical coordinate, and the magnitude of the received power as the intensity change of the two-dimensional image. The actual channel propagation may experience two or more times reflections, which is difficult to realize in the modeling, due to the limited information in the measurement by our setup. So the THz channel model is established by considering specular reflection based on the reflection characteristics of the wall and window surfaces. The signal path loss includes two parts: the free space propagation loss (FSPL) and the reflection loss. The reflection loss is caused by the reflection of the signal through the building walls or windows, which could be obtained by the measured reflectivity (Fig. 3(b,e)) and also by the calculated reflectivity (Fig. 3(c,f)) based on perceived refractive indices. According to the established THz channel model, the spatial power distribution map is made in the same way. The results are shown in Fig. 3 for the channel measurement and modeling results. It can be found that the spatial power distribution derived from the material reflectivity measured using the experiment and that calculated using the theory is basically the same, which proves the accuracy of the above method of measuring the material reflection characteristics. Compare spatial power distributions obtained by measuring and sensing material properties. Meanwhile, from the overall modeling results, it can be seen that the spatial power distribution of channel measurement and channel modeling are basically the same, indicating that the channel model is reliable. Moreover, it can be seen from Fig. 3 that the received power is large when propagating through LOS path, indicating that the signal transmission quality is high. In addition, when propagating through the NLOS path, the signal transmission through the reflection of the left building has a higher signal transmission quality than the signal transmission through the reflection of the right building. Therefore, the transmission quality of the signal can be evaluated by the received power. However, the spatial power distribution of the channel measurement is more scattered, because there are two or more times reflections during the channel measurement process due to the shape of the building [36], which should be recognized by measurement of time delay. In our channel models, only the LOS and specular reflection components are considered, for the concern that two or more times reflections can lead to higher reflection loss and free space propagation loss [37]. This makes the two or more times reflection components more attenuated and difficult to be collected by THz receivers in actual scenarios [38].

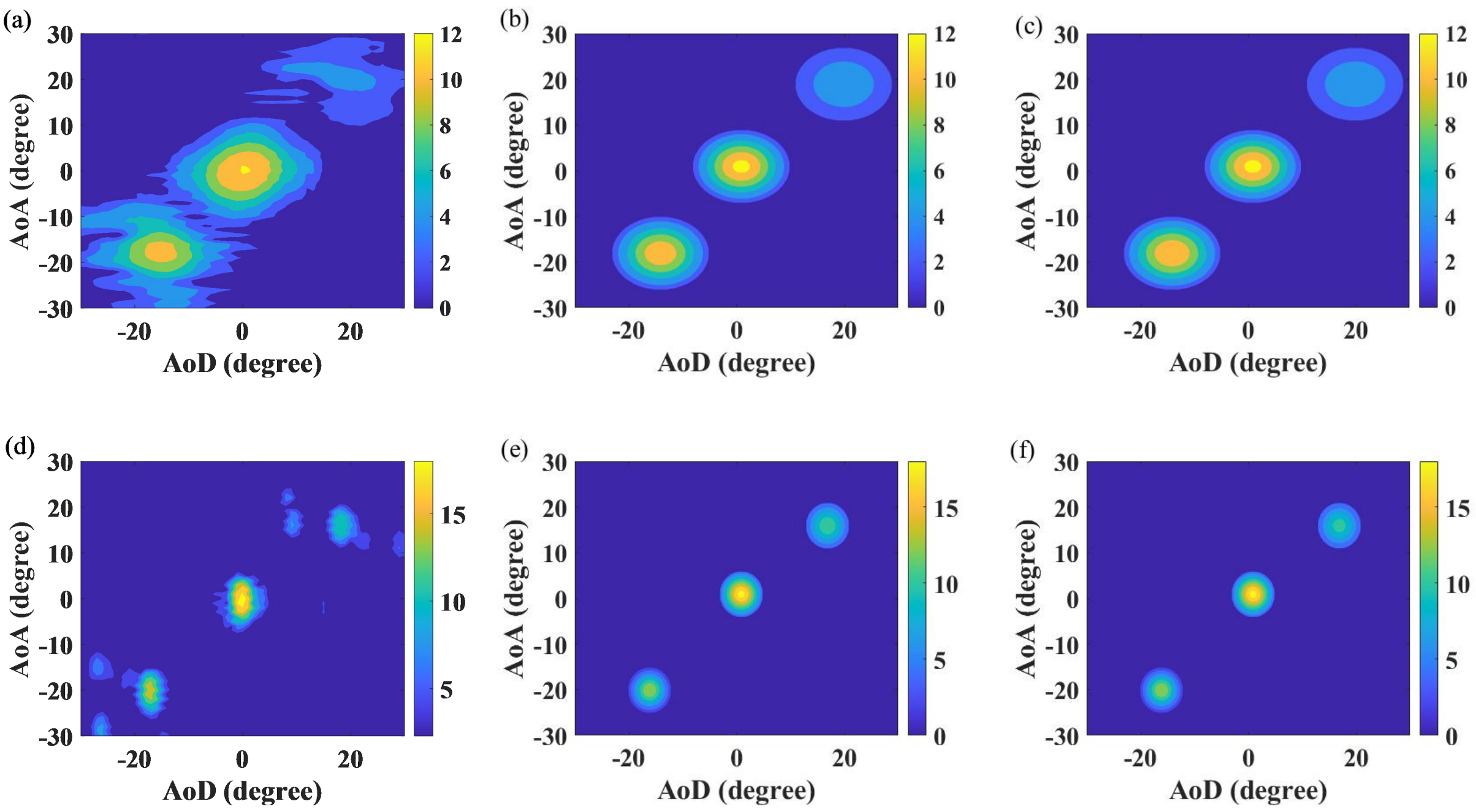

**Figure 3**. Channel measurement and modeling. (a) Channel measurement at 140 GHz; (b) Channel modeling based on measured reflectivity at 140 GHz; (c) Channel modeling results based on predicted reflectivity at 140 GHz. (d) Channel measurement at 240 GHz; (e) Channel modeling based on measured reflectivity at 240 GHz; (f) Channel modeling results based on predicted reflectivity at 240 GHz.

Since the transmitted signal propagates through multiple paths created by reflections from the building and the length of each path is different, this results in the path components arriving at the Rx at different times. This causes the signal to exhibit delay expansion in the time domain, i.e., the signal received by the Rx contains MPCs with different delays and different magnitudes [39]. The signal is presented at the receiving end as a superposition of multiple incidence angles, which forms an angular expansion that gives the received signal a wide angular distribution. At the same time, because the signals on different paths are reflected by the building model to different degrees during the propagation process, they will experience different degrees of attenuation, which will lead to spatial fading of the signal intensity in the airspace. Signals on different paths may interfere or superimpose each other in space, forming spatial correlation [20, 40]. Therefore, based on the measured data, the channel characteristics at 140 GHz and 240 GHz are analyzed as shown in Fig. 4 by showing the CDF of path delay to ~~characterize~~ describe the characteristics of THz channel. The difference between the delays at the beginning is due to the different LOS channel distances (between both antennas at the Tx and Rx sides). It can be seen that the number of LOS paths of the 140 GHz channel is basically below 40%, which indicates that there are more NLOS paths due to the shading of the buildings on both sides. Meanwhile, due to the different geometries of the buildings, the channel propagation process will have 2 or more times of propagation, generating MPCs, which makes the path delay different [41]. However, there are more LOS paths for the 240 GHz channel due to the narrower of its beam width (4°). Such a narrower beam width results in a more focused signal that is less prone to spreading or bending around obstacles. Besides, at higher frequencies, the channel beam becomes more focused, which can enhance directional communication but also makes the channel more susceptible to obstruction and thus more distinct in LOS versus NLOS conditions [42].

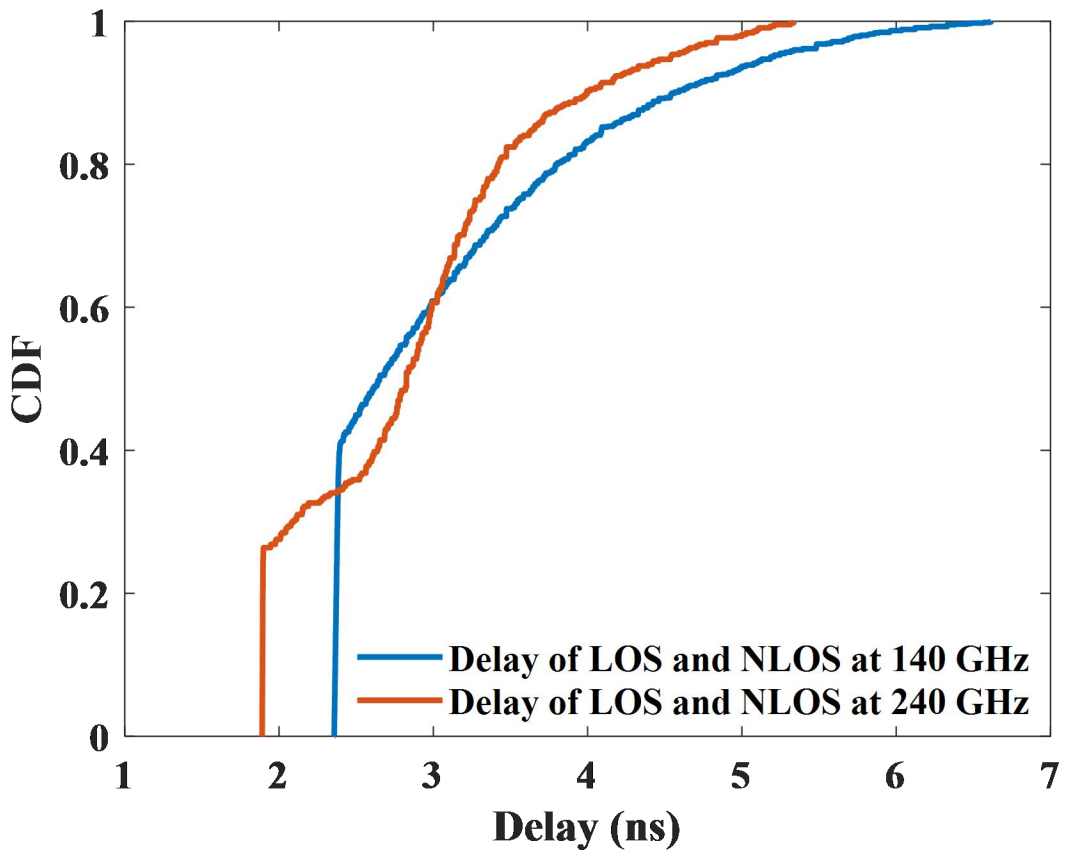

**Figure 4**. CDF of time delay performance.

### IV. Channel measurement and modeling in indoor scenario

The miniaturized setup alone may not fully demonstrate the efficiency of this method. To account for potential differences between the miniaturized setup and real-world environments, we also conducted indoor experimental measurements to validate the scalability and correctness of our results. We perform actual indoor scenario channel measurements (Fig. 5(a)) and modeling, and compare the results. Channel measurements are performed in an indoor scenario using the equipment described above with the channel frequency set at 140 GHz. In this measurement, we keep the azimuth angle to be 0° always and varied the elevation angles of the Tx and Rx thus characterizing the actual reflections by the ceiling and the floor. The experimental validation is carried out in a conference room with a total length of 8 m and the ceiling height of 2.71 m. The height of Tx and Rx is 0.86 m, the distance from both sides of the wall is 1.5 m, and the distance between Tx and Rx is 5 m. Both the Tx and Rx units are mounted on separate rotary tables. Setting the rotary table under the Tx to rotate from -18° to +40° and the rotary table under the Rx to rotate from -40° to +60°. The rotation step is still 1° again. The Tx and Rx are positioned on opposite sides of the indoor scenario, as shown in Fig. 5(a). By changing the elevation angles of Tx and Rx, we make multiple reflection measurements of ceiling and floor to perceive their dielectric properties.

The measured spatial power distribution results are shown in Fig. 5(b). It can be seen that the NLOS path is generated due to reflection by the floor when the elevation angle is less than 0°, and the NLOS path is generated due to reflection by the ceiling when the elevation angle is larger than 0°. It can be found that the power generated by the floor's reflection is greater than that by the ceiling's reflection through the channel measurements. At the same time, because the ceiling and the floor are smooth, it can be seen that the spatial is less dispersion in the power distribution. From Fig. 5(c), it can be seen that using the perceived refractive indices (1.998 for floor and 1.621 for ceiling by Eq. (2)) for channel modeling, the modeling results are basically the same as the channel measurement results [43]. Because the specular reflection is taken into account in the modelling, while the floor and ceiling have a certain degree of roughness in the actual environment, so there is a slight difference in the power distribution. Such a difference can be further reduced by introducing a roughness factor into the theoretical model [37] and makes it more accurate in future channel modeling. It should be noted that the refractive indices can be perceived by using Eq. (2) when there is no multiple reflection components [37] and the perceived value for the plater ceiling is close to the measured one by THz-TDS [44]. This helps to corroborate the findings obtained from the miniaturized setup, providing further confidence in the scalability and accuracy of the proposed approach in real-world environments.

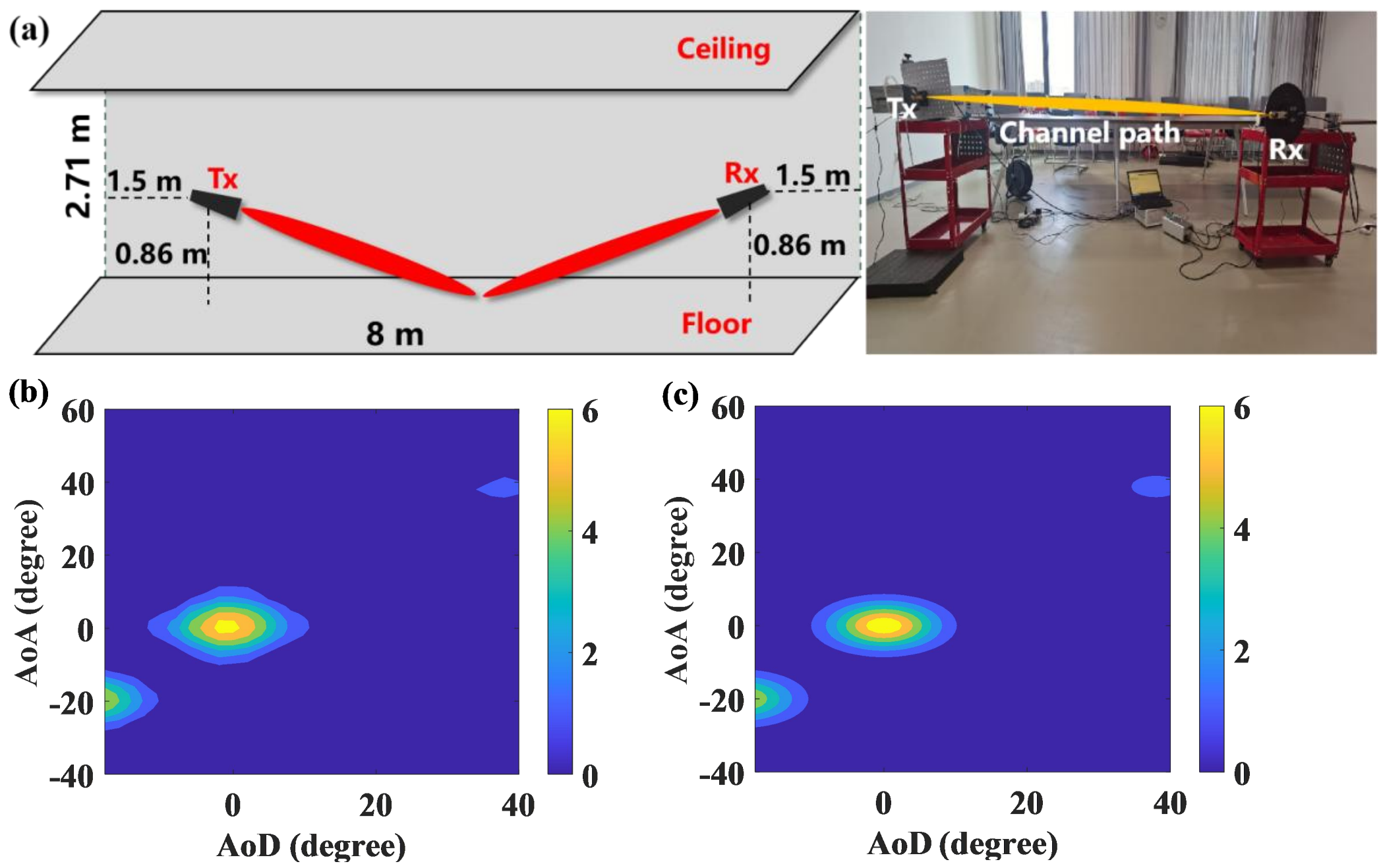

**Figure 5**. 140 GHz indoor channel measurement and simulation results. (a) Configuration of indoor channel measurement; (b) Measured spatial power distribution; (c) Modelling spatial power distribution;.

Based on the indoor channel measurements, we still characterize the indoor channel in terms of received power and delay extension. The MPCs in the indoor scenario are generated primarily due to reflections from the surfaces of the ceiling and the floor. These reflections cause scattering of the THz signals, resulting in the observed MPCs. The results are shown in Fig. 6. It can be seen that the indoor scenario channel measurements are mainly LOS paths, and less NLOS is generated due to the ceiling and floor reflections, i.e., the MPCs are lower. The power of the LOS paths is higher than that of the NLOS paths reflected by the floor and the ceiling. Compared to the path delay due to reflection by the ceiling, the path delay due to the floor is more dispersed, contributing to the majority of MPCs [43]. This is due to the smaller reflection loss (-5.5 dB by the floor and -8.8 dB by the ceiling) and FSPL (longer distance to the ceiling than to the floor), which introduce higher power attenuation into the NLOS paths due to reflection by the ceiling.

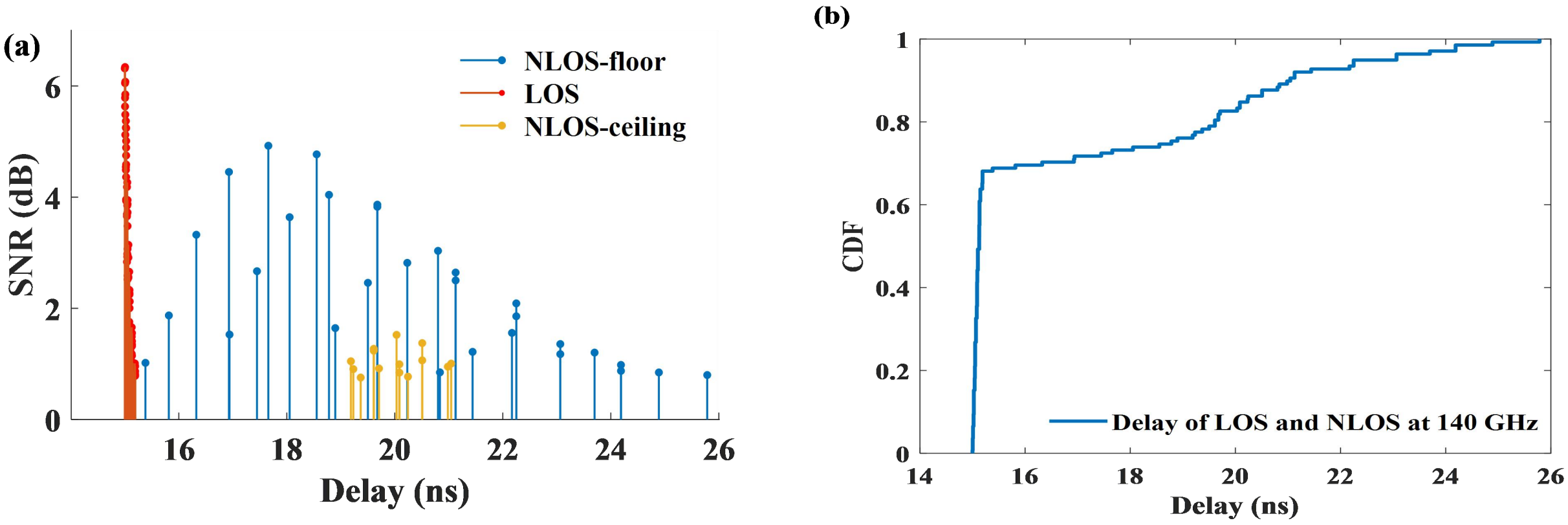

**Figure 6.** Analysis of 140 GHz indoor channel characteristics. (a) Time domain characteristic analysis; (b) CDF of time delay performance.

Therefore, no matter in any unknown environment, the above method can be used to deduce the exact dielectric properties, so as to establish an accurate channel model.

## V. Conclusion

For THz channel modeling, particularly within real-world settings, the dielectric characteristics of environmental surfaces are pivotal for achieving precise models. These properties have traditionally been determined through direct measurements using THz-TDS or VNA (vector network analyzers). However, such measurements may not be viable in environments that have not been previously characterized. In this work, we introduce an innovative channel modeling method that leverages the natural sensing abilities of THz channels by using Fresnel equations, assessing its efficacy within both a miniaturized cityscape scenario and an indoor scenario. Our results demonstrate that the values perceived by our THz channels closely align with those obtained through THz-TDS measurements, exhibiting errors below 6%. This finding underscores the potential of our approach to accurately sense and deduce the dielectric properties of surfaces in scenarios where traditional characterization is not possible. Utilizing a ray-tracing algorithm, we find that the spatial power distribution derived from channel modeling with the perceived refractive indices closely matches the distribution from direct channel measurements, further validating the precision of our proposed method. Moreover, this method facilitates the analysis of MPCs by simulating path time delays. However, it’s important to acknowledge that the presence of even minimal errors suggests there is scope for enhancing our methodology, possibly by refining the theoretical model to account for factors like surface roughness. This method should be investigated in more complicated scenarios with rough surfaces in future to demonstrate its efficiency further.